\documentclass[5p,times,twocolumn]{elsarticle}
\usepackage{graphicx}
\usepackage{dcolumn}
\usepackage{bm}
\usepackage{multirow}
\usepackage{tablefootnote}
\usepackage{color}

\begin{document}

\begin{frontmatter}

\title{Modification of Nanodiamonds by Xenon Implantation: A Molecular Dynamics Study}

\author[curtin]{Jason L. Fogg}
\author[curtin]{Alireza Aghajamali}
\author[UK]{Jonathan A. Hinks}
\author[UK]{Stephen E. Donnelly}
\author[Russia]{Andrey. A. Shiryaev}
\author[curtin]{Nigel A. Marks \corref{cor1}}
\cortext[cor1]{Corresponding author. N.Marks@curtin.edu.au}

\address[curtin]{Department of Physics and Astronomy, Curtin University,  Perth WA 6102, Australia}
\address[UK]{University of Huddersfield, Queensgate, Huddersfield, HD1 3DH, United Kingdom}
\address[Russia]{Frumkin Institute of Physical Chemistry and Electrochemistry RAS, Leninsky pr .31 korp. 4, Moscow, 119071, Russia}

\begin{abstract} 
Xenon implantation into nanodiamonds is studied using molecular dynamics. The
nanodiamonds range in size from 2--10~nm and the primary knock-on (PKA) energy
extends up to 40~keV.  For small nanodiamonds an
energy-window effect occurs in which PKA energies of around 6~keV destroy the
nanodiamond, while in larger nanodiamonds the radiation cascade is increasingly
similar to those in bulk material. Destruction of the small nanodiamonds occurs
due to thermal annealing associated with the small size of the particles and
the absence of a heat-loss path. Simulations are also performed for a range of
impact parameters, and for a series of double-nanodiamond systems in which a
heat-loss path is present.  The latter show that the thermal shock caused by
the impact occurs on the timescale of a few picoseconds.  These findings are
relevant to ion-beam modification of nanoparticles by noble gases as well as
meteoritic studies where implantation is proposed as the mechanism for xenon
incorporation in pre-solar nanodiamonds.
\end{abstract}


\end{frontmatter}

\section{Introduction}

Nanodiamonds are found in primitive chondritic meteorites at concentrations as
high as 1500~ppm \cite{Huss-Elements-2005}. Within these nanodiamonds (NDs) are trace amounts of xenon
and other noble gases whose unusual isotopic abundances indicate a pre-solar
origin and provide information on nucleosynthesis processes in supernovae.
Ion-implantation is the most popular explanation for the presence of the
noble gases in NDs \cite{Anders-Meteoritics-1993,Lewis-Nature-1987,Verchovsky-Science-1998,Daulton-GCA-1999}, 
although another possibility is that the
noble gases were incorporated during growth of the ND. The implantation
hypothesis was studied experimentally by Koscheev et al. \cite{Koscheev-Nature-2001} and Verchovsky et
al. \cite{Verchovsky-Abs-2000} using low-energy implantation into detonation nanodiamonds, which are
similar in size to meteoritic NDs. Even though their studies employed only a
single implantation energy and used isotopes of natural abundance, many of the
characteristics observed in the meteoritic data were observed.

Recently we performed a combined experimental and computational study of xenon
implantation into nanodiamonds of varying size \cite{Shiryaev-SciRep-2018}. Implantation
experiments were performed \emph{in situ} in a transmission electron
microscope (TEM), enabling real-time monitoring of irradiation effects. While large
NDs (40~nm diameter) were resistant to the ion beam, the small NDs (2--3~nm) were
gradually destroyed under 6~keV irradiation.  SRIM calculations do not predict
this effect, while Molecular Dynamics (MD) simulations explained these
observations via a temperature effect, which anneals the smaller-sized NDs. The
success of the MD approach is that it goes beyond the binary-collision
approximation used by SRIM, and captures all the thermal motion associated with
the collision cascade.

In this work our MD studies of xenon implantation into ND is extended in two main
ways: (i) low implantation energies are considered to study the hypothesis
explored by Koscheev et al. \cite{Koscheev-Nature-2001} and Verchovsky et al. \cite{Verchovsky-Abs-2000}, and (ii) a second ND
particle is introduced to capture thermal conduction effects relevant to the
\emph{in situ} experiments. We also examine the effect of varying the impact parameter and include a comprehensive description of the
simulation approach, including the description of the Xe--C interactions and the
methodology for generating the ND coordinates. The latter is a useful starting point for further investigations of noble gases within nanodiamonds.

The simulations are performed using the environment dependent interaction
potential \cite{Marks-PRB-2000,Marks-JPCM-2002} which provides an excellent description of both
$sp^2$ (graphite) and $sp^3$ (diamond) bonding \cite{deTomas-Carbon-2016}. A large number of
primary knock-on atom (PKA) directions and energies are used to collect a
statistically significant data set. The nanodiamonds vary in size from 2 and
10~nm, and PKA energies from 50~eV up to 40~keV are considered.

\section{Methodology}

Atomic interactions between carbon atoms were described using the
environment dependent interaction potential (EDIP) for carbon,
coupled with the Ziegler-Biersack-Littmark (ZBL) potential \cite{ZBL-1985}
for close interactions. This combination has been previously been
successfully employed to model radiation damage cascades in
graphite \cite{Christie-Carbon-2015, Vukovic-PRApplied-2018} and diamond \cite{Buchan-JAP-2015}.  Full details
of the interpolation process used to switch between the EDIP and
ZBL forms is provided in Ref.~\cite{Christie-Carbon-2015}. The EDIP functional
form provides an excellent description of the competing hybridizations
in carbon, in particular the energy barrier between graphite and
diamond. In a recent comparison of carbon potentials \cite{deTomas-Carbon-2016}, 
we found that EDIP has excellent transferability across a wide range of 
conditions, and is superior to many common carbon potentials. One
aspect of EDIP which is still being developed is the ability to 
describe hydrogen, and hence all nanodiamonds in the simulations
are dehydrogenated.

\begin{figure}[t]
\centering
\includegraphics[width=\columnwidth]{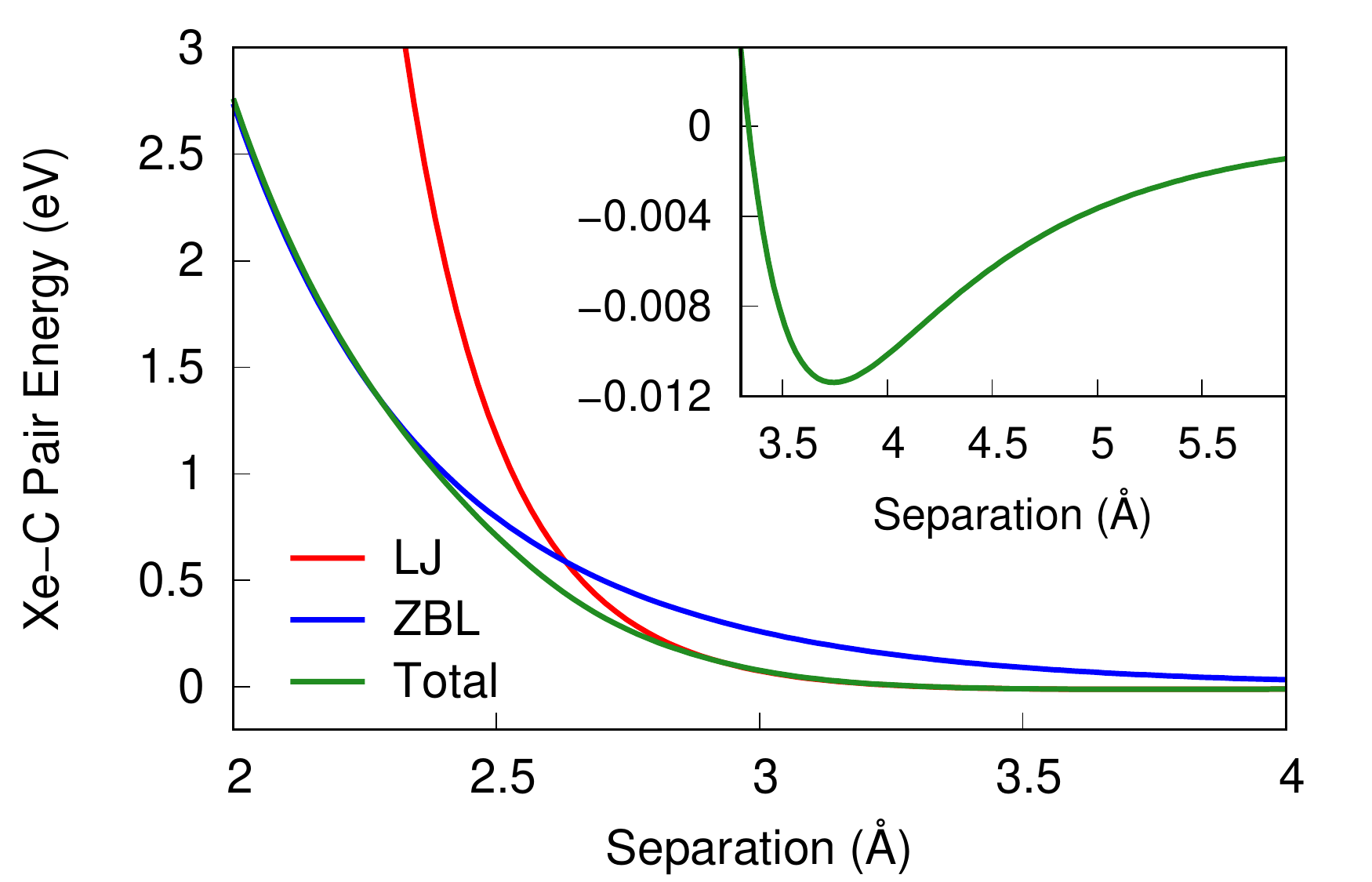}
\caption{ Interaction energy between the Xe and carbon atoms (green line). At
close approach the interaction is pure ZBL (blue line), while at distances
around equilibrium and greater a Lennard-Jones (LJ) expression is used (red line). 
Interpolation at intermediate distances is performed using a Fermi-like switching 
function as described in the text. }
\label{potential}
\end{figure}

For the Xe-C interactions, we used the standard ZBL potential coupled with
a Lennard-Jones potential of the form
\[
    U(r) = 4 \varepsilon_\mathrm{Xe-C} 
             \left\{
                \left( \frac{\sigma_\mathrm{Xe-C}}{r} \right)^{12} - 
                \left( \frac{\sigma_\mathrm{Xe-C}}{r} \right)^6
             \right\}
\]
with parameters $\varepsilon_\mathrm{Xe-C}$=0.0114~eV and
$\sigma_\mathrm{Xe-C}$=3.332~\AA\ as given in Ref.~\cite{Simonyan-JCP-2001}.
For simplicity, we did not employ the anistropic terms as implemented in
Ref.~\cite{Simonyan-JCP-2001}. Interpolation between the LJ form and the
ZBL interaction was described in the same manner as Refs.~\cite{Christie-Carbon-2015,Buchan-JAP-2015},
using Fermi-type switching functions. The parameters were empirically determined
to provide a smooth transition from the strong repulsive region of the ZBL to the
weakly attractive region of the LJ potential. The resulting functional form of the 
energy for Xe-C interactions was
\[
    U(r) = U_\mathrm{ZBL}(r) \times    f(r+\delta) + 
           U_\mathrm{LJ}(r)  \times (1-f(r-\delta))
\]
where $\delta$=0.07~\AA\ and
\[
    f(r) = \big[ 1 + \exp \big(8\times(r-2.7)\big) \big]^{-1}.
\]
A plot of the Xe-C interaction energy covering the ZBL and LJ regimes is shown
in Fig.~\ref{potential}, with the attractive region shown in the inset.

The simulations were performed using an in-house Molecular Dynamics package
developed by one of the authors (NAM). All calculations were performed in an
NVE ensemble (constant number of particles, volume and energy), using Verlet
integration and a variable timestep \cite{Marks-NIMB-2015}. The algorithm for
the timestep automatically adjusts the timestep using the metric
$||\mathbf{F}_\mathrm{max}||\Delta t$ as described in detail in
Ref.~\cite{Marks-NIMB-2015}.  During close approaches the timestep was reduced
to values as small as $10^{-5}$~fs, and was subsequently increased as kinetic
energy was dispersed and the system equilibrated.  Periodic boundary conditions
were not employed. Prior to implantation, all nanodiamond coordinates were relaxed
by steepest descent and equilibrated at 300~K. The majority of the
simulations were 1~ps in length, which was sufficient to model the ballistic
phase of the Xe implantation onto the nanodiamond. A smaller number of longer
simulations extending up to 85~ps were performed to study annealing effects and
thermal transport.  Visualization was performed using the OVITO software
\cite{Stukowski-MSMSE-2010} and a cutoff of 1.85~\AA\ was to determine the coordination
number.  Temperatures were determined using the kinetic energy of the main
carbon cluster; the xenon atom and any ejected carbon atoms were
excluded from the calculation and the net momentum of the cluster was subtracted
prior to computing the temperature.

\begin{table}[b]
\caption{ Parameters (number of atoms and distances), diameter and $sp^3$ fraction for the twelve NDs studied in this work. }
\begin{center}
\begin{tabular}{l l l l l}
Diameter~(nm) & N$_\mathrm{atoms}~~~~$ & $d_{111}$ (\AA) & $d_{100}$ (\AA) & $sp^3$ (\%) \\
\hline\hline              
2.1 &  837  &    9   & 10 & 69.9 \\ 
2.6 & 1639  &   12   & 14 & 74.9 \\ 
3.1 & 2793  &   14   & 17 & 78.4 \\ 
3.6 & 4363  &   15   & 21 & 81.0 \\ 
4.0 & 5975  &   18   & 21 & 83.7 \\ 
4.6 & 8389  &   20   & 24 & 85.4 \\ 
5.1 & 11591  &  22   & 27 & 86.6 \\ 
6.0 & 18977  &  26   & 31 & 88.8\\ 
7.1 & 29359  &  30.5 & 35.5 & 90.4\\ 
8.0 & 46393  &  35   & 46 & 91.3 \\ 
9.0 & 61849  &  39   & 46 & 92.4 \\ 
10.2 & 90395  & 44   & 49 & 93.5
\label{table}
\end{tabular}
\end{center}
\end{table} 

Construction of the nanodiamond coordinates is a non-trivial problem. The
question of stable nanodiamond morphologies has been extensively studied by
Barnard and Zapol, \cite{Barnard-JCP-2004} and some sets of Cartesian coordinates are
available online \cite{Barnard-CSIRO-2014}. For dehydrogenated nanodiamonds, they showed that
the relevant stable geometry is the truncated octahedron, formed by cleaving a
carbon nanoparticle out of diamond such that it has only 100 and 111 faces (see
Fig.~8 in Ref~\cite{Barnard-JCP-2004}). One important detail of dehydrogenated
nanodiamonds is that the 100 surfaces need to be reconstructed in a $2\times1$
arrangement to eliminate dangling bonds (two-fold coordinated atoms) on the 100
surface. All of these aspects were implemented in a Fortran program which is
available on request.  The program generates a nanodiamond of arbitrary size,
with all atoms at the surface being $sp^2$ bonded (the 100 surface due to the
reconstruction, and the 111 surface by design). Table~\ref{table} lists all of 
the NDs studied in this work; the parameters $d_{100}$ and $d_{111}$ are the 
distances from the origin to the 100 and 111 planes, respectively.
The radius was measured by computing the average distance between $sp^2$
bonded atoms and the centre-of-mass of the ND.
The NDs studied in this work span the typical range observed for meteoritic ND (range of 1--10~nm, with average diameter of around 3~nm; Ref.~\cite{Dai-Nature-2002,Huss-Elements-2005}) and detonation ND (average of 4--5~nm). Our largest ND of 10~nm is of sufficient size to exhibit bulk-like behaviour, and hence we did not simulate the 40~nm diamonds studied in Ref.~\cite{Shiryaev-SciRep-2018}.

\begin{figure}[t]
\centering
\includegraphics[width=0.85\columnwidth]{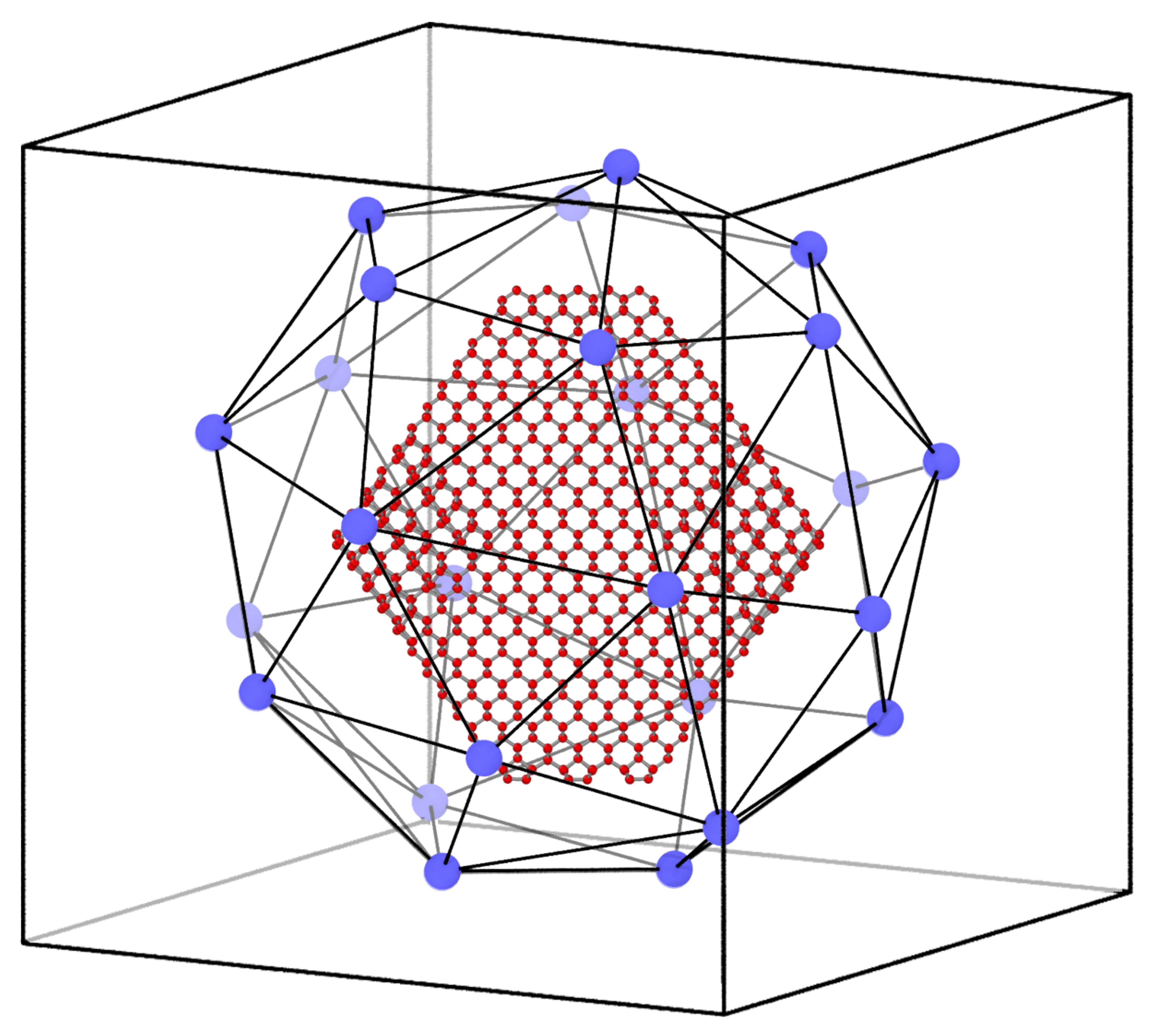}
\caption{ Schematic showing the typical geometric relationship between a ND 
and the initial PKA locations. The ND atoms are shown in red, and in this
example a 4.0~nm diameter ND is shown. The PKA locations (shown in blue) are solutions 
to the 25-point Thomson problem. When the PKA is initiated, the Xe travels towards 
the centre of the ND. For calculations studying the effect of impact parameter, the 
ND is offset as described in the text. }

\label{thomson}
\end{figure}

The energy of the xenon PKA was varied between 2 and 40~keV, with the majority
of simulations performed using 6~keV.  The initial position of the xenon was
always 5~\AA\ beyond the edge of the ND and for most simulations the
initial velocity vector was towards the centre of the ND. In a small
number of simulations (see Supplementary Material) the ND was displaced perpendicular to the
initial velocity vector to study the effect of varying the impact parameter. The result was
as intuitively expected; namely that a higher impact parameter progressively 
transfers less energy to the ND, tending to zero effect when the Xe grazes the
edge of the ND.
To collect robust statistics across the various crystallographic orientations,
the xenon atom was introduced at 25 different locations distributed uniformly
around the ND. A schematic indicating a ND and the various
xenon positions is shown in Fig.~\ref{thomson} and Supplementary Movie~S1; note that the location of the
xenon atoms has been expanded outwards for clarity. The 25 points shown are
solutions to the Thomson problem \cite{Thomson-1906}, an exercise in
mathematical optimization in which point charges repel each other on the unit
sphere. For further examples of the usefulness of the Thomson problem and its
applicability to radiation damage simulations, see
Refs.~\cite{Christie-Carbon-2015,Buchan-JAP-2015,Robinson-PRB-2012-86,Robinson-PRB-2012-85,Robinson-MCP-2014}.

\section{Results}

\begin{figure}[b]
\centering
\includegraphics[width=\columnwidth]{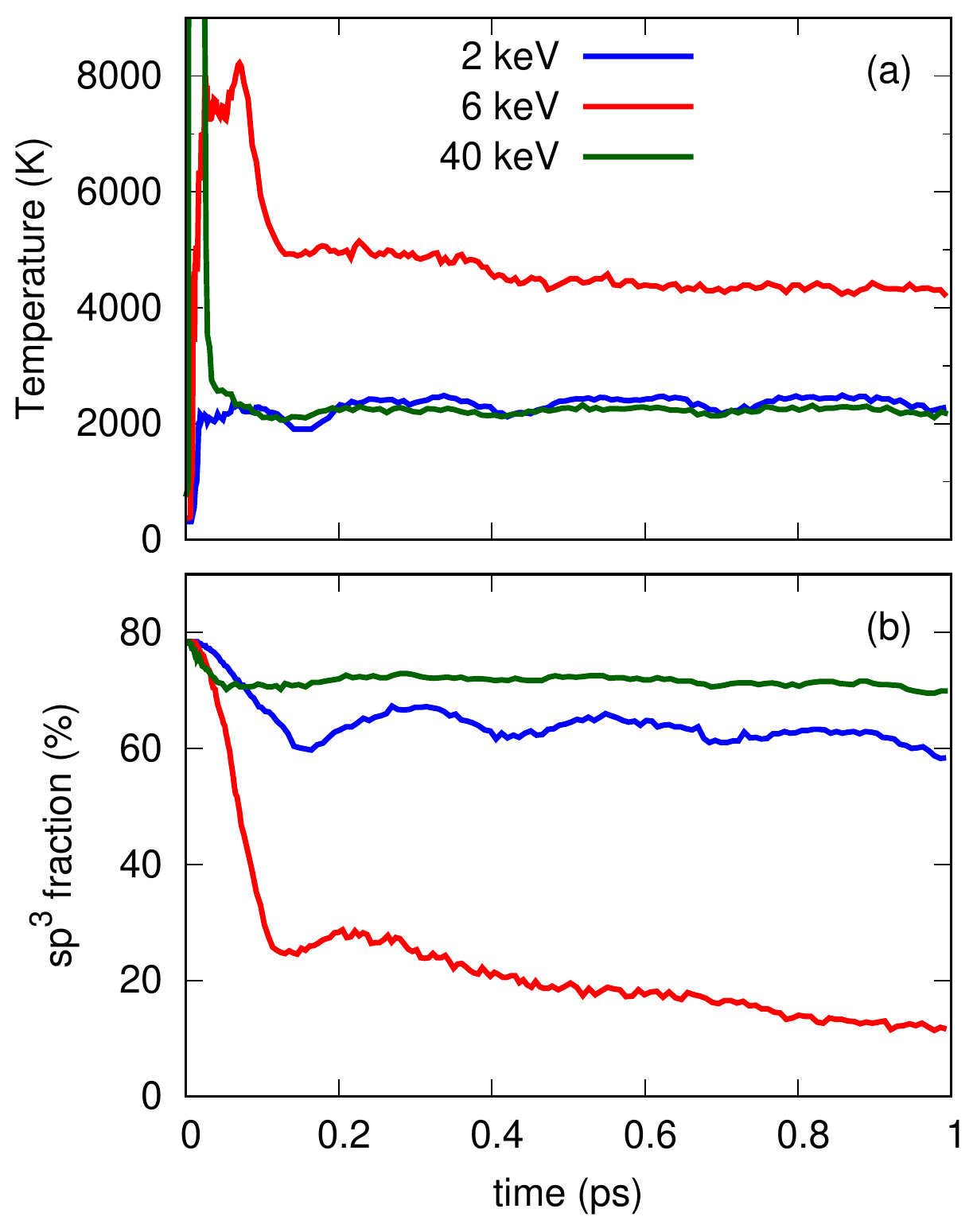}
\caption{Time evolution of the cluster temperature [panel (a)] and $sp^3$ fraction
[panel (b)] for the three xenon implantation events shown in Fig.~\ref{time-evolution-images}. }
\label{time-evolution}
\end{figure}

\begin{figure*}[t]
\centering
\includegraphics[width=2.05\columnwidth]{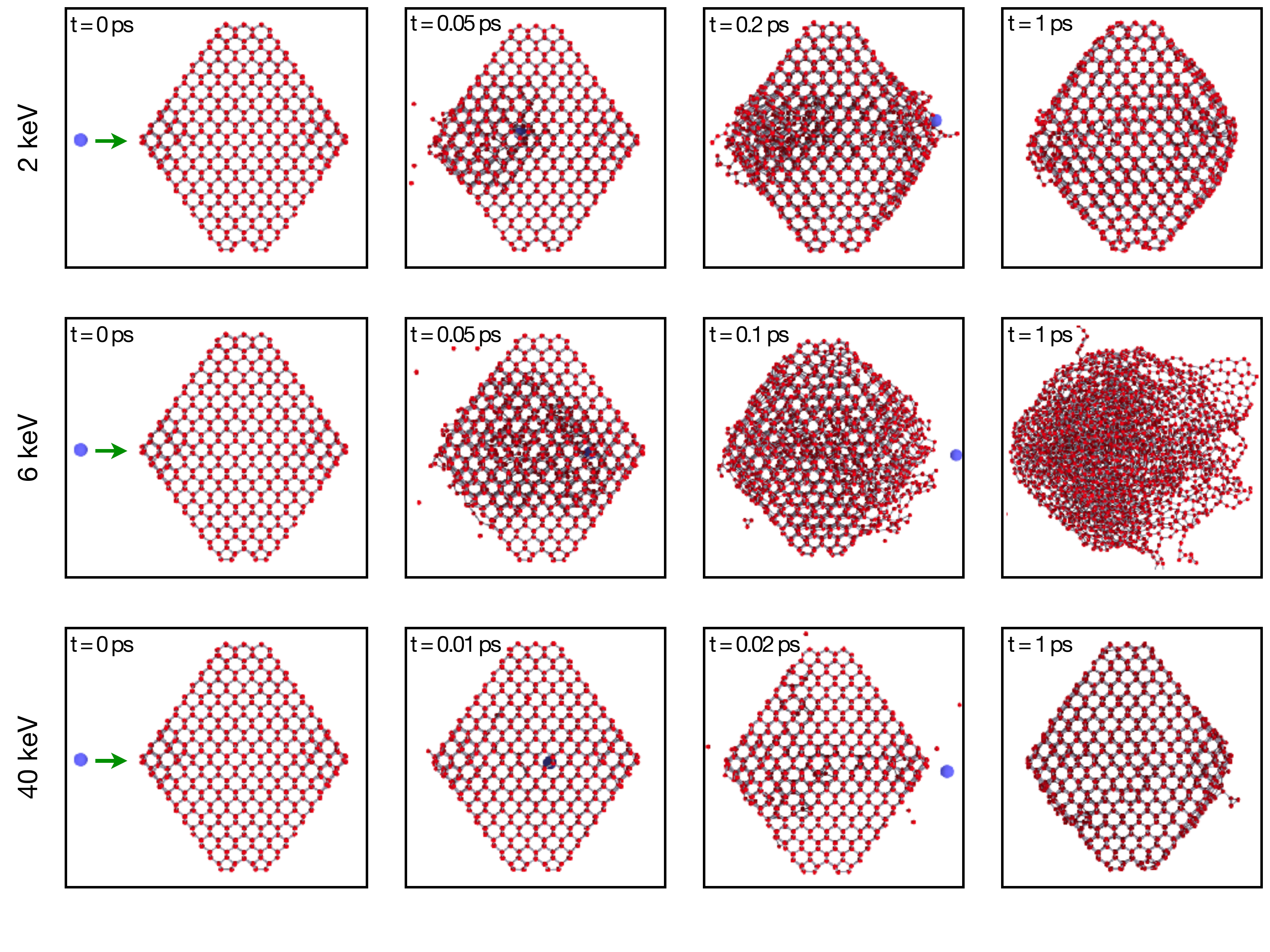}
\caption{Time evolution of a three typical implantation events into a 3.1~nm
diameter nanodiamond for PKA energies of 2, 6 and 40~keV. In each case the Xe
PKA exits the nanodiamond. Only at 6~keV is there any significant destruction
of the nanodiamond. Carbon atoms and the xenon are shown as red and blue circles
respectively. }
\label{time-evolution-images}
\end{figure*}

The implantation of a xenon onto the ND is a highly energetic event,
leading to substantial disruption in certain cases.
Figs.~\ref{time-evolution} and \ref{time-evolution-images} show 
representative implantations for varying PKA energies incident onto a 3.1~nm
diameter ND. In each case the initial position and
velocity vector is the same; Fig.~\ref{time-evolution} plots the time-variation of the
cluster temperature and $sp^3$ fraction, while Fig.~\ref{time-evolution-images} shows a series of snapshots. Video animation sequences of all three
impacts are provided in the Supplementary Material.

At 2~keV the Xe atom deposits almost all of its kinetic energy into the
ND. As shown in the top panel of Fig.~\ref{time-evolution-images}, the
Xe takes 0.2~ps to travel through the ND, ejecting a small
number of carbon atoms (six) in the process.  For the case shown, the Xe exits
the ND with just 24~eV of energy, while for other PKA directions the Xe
end-of-range sometimes falls within the ND. The transfer of kinetic
energy into the ND rapidly increases the temperature of the ND, as shown in
Fig.~\ref{time-evolution}(a); after a very short period (less than 0.1~ps) the
temperature reaches equilibrium at a value slightly above 2000~K. This
temperature is sufficient to slightly reduce the number of $sp^3$ bonds
[Fig.~\ref{time-evolution}(b)] in the ND, principally at the 111 edges on the
right-hand side of the ND where a small amount of graphitization occurs.
Fig.~\ref{time-evolution}(b) also shows that for the 2~keV impact the $sp^3$
fraction undergoes oscillations which gradually die away; these reflect a
``ringing'' of the structure in which sound waves reflect back-and-forth across
the ND.

The bottom panel of Fig.~\ref{time-evolution-images} shows that implantation
of a 40~keV Xe has a similar effect on the ND as seen at 2~keV. The chief 
difference is that the Xe passes through the ND extremely quickly, taking 
around 0.02~ps as seen in the figure. The Xe exits the ND with a very high
kinetic energy (23.2~keV), and also ejects a significant number of carbons 
(23 atoms). The combined kinetic of the ejected carbons is 15.0~keV, meaning
that only 1.8~keV of the original Xe PKA energy is deposited into the ND.
This results in an equilibrium ND temperature of around 2000~K as shown in 
Fig.~\ref{time-evolution}(a). Since the temperature is slightly less than
for the 2~keV case, the annealing effect which reduces the $sp^3$ fraction
is also less significant, as can be seen in Fig.~\ref{time-evolution}(b).

The behaviour at 6~keV is completely different to the higher and lower
energy cases. Here, a much larger amount of kinetic energy is transferred
to the ND. As shown in the sequence in the middle row of 
Fig.~\ref{time-evolution-images}, the Xe exits the ND after 0.1~ps, but even
at this stage there is already a significant amount of structural damage to
the ND. The final kinetic energy of the Xe is 0.4~keV, while a total of
67 carbon atoms are ejected, with a combined kinetic energy of nearly 1~keV.
In total, around 4.7~keV of kinetic energy is transferred to the ND, increasing
the instantaneous temperature to well above the melting point (circa 4300--4500~K for EDIP) 
as shown in Fig.~\ref{time-evolution}(a). Even once the temperature has 
equilibrated, the temperature is extremely high, at just over 4000~K. At such a high
temperature there is substantial conversion of $sp^3$ into $sp^2$ bonding
in combination with some evaporation of atoms. As seen in Fig.~\ref{time-evolution}(b),
the $sp^3$ fraction is greatly reduced, and after 1~ps the majority of the tetrahedral
bonding is lost.

\begin{figure}[t]
\centering
\includegraphics[width=\columnwidth]{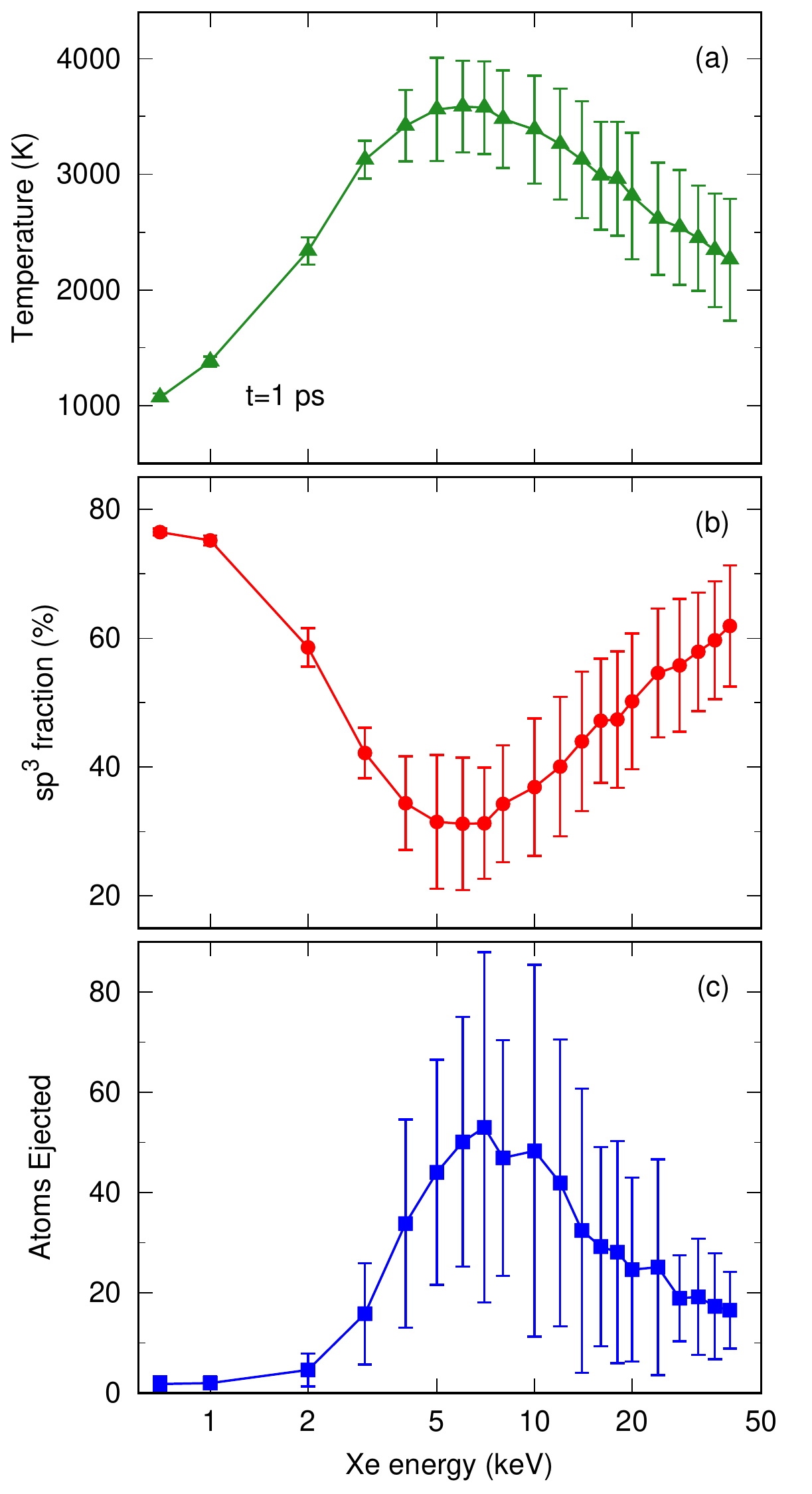}
\caption{ Effect of PKA energy for Xe implantation into a 3.1~nm diameter 
ND.  All properties computed 1~ps after initiation of the PKA.
(a) Cluster temperature, (b) $sp^3$ fraction, and (c) number of carbon atoms ejected. }
\label{PKAenergy-3.1nm}
\end{figure}

Ion implantation processes are sensitive to the impact parameter of the primary collision and
crystallographic orientation, and hence it is necessary to average over many
different directions to collect accurate statistics.  In the case of the 6~keV
impact, for example, we found that for our 25 initial directions the final
temperature of the ND varied from a low of 2520~K to a high of 4160~K. Similar
variability was observed for all other quantities we extracted from the data.
The merits of averaging are seen in Fig.~\ref{PKAenergy-3.1nm} which shows
the variation in the temperature, $sp^3$ fraction and number of carbon atoms
ejected as a function of Xe energy. The error bars indicate the standard deviation, 
and even though the spread is substantial, the lines linking
the mean values themselves are quite smooth, particularly for the temperature
and $sp^3$ fraction. When the Xe energy is around 6~keV damage to the ND is
maximal, with the sp$^3$ fraction [Fig.~\ref{PKAenergy-3.1nm}(b)] reduced to
well under half its initial value and around 50 carbon atoms being ejected
[Fig.~\ref{PKAenergy-3.1nm}(c)]. For higher Xe energies the nuclear stopping
efficiency decreases significantly, leading to a lower residual temperature and
much less damage (as measured by the $sp^3$ fraction and number of atoms
ejected). At the highest energy considered, 40~keV, the effect of the Xe is
very similar to that at 2~keV, consistent with the single impact sequences
shown in Fig.~\ref{time-evolution-images}.

The energy dependence of all three quantities in Fig.~\ref{PKAenergy-3.1nm} are
closely correlated, highlighting that the heating of
the ND is the critical quantity that drives the damage process. When the ND
temperature is 1000~K or so the $sp^3$ bonds are able to withstand the
implantation due to the high melting point of diamond. However, when the
temperature approaches several thousand degrees, $sp^3$ bonded atoms change to
$sp^2$ and the ND begins to convert to a nested fullerene structure, or carbon
onion, as shown in Fig.~\ref{6keV-snapshots}(a).  The large number of $sp^2$
bonded atoms is seen in the large number of green atoms, consistent with the
$sp^3$ fraction of $\sim$30\% at 6~keV as seen in Fig.~\ref{PKAenergy-3.1nm}.

\begin{figure}[b]
\centering
\includegraphics[width=\columnwidth]{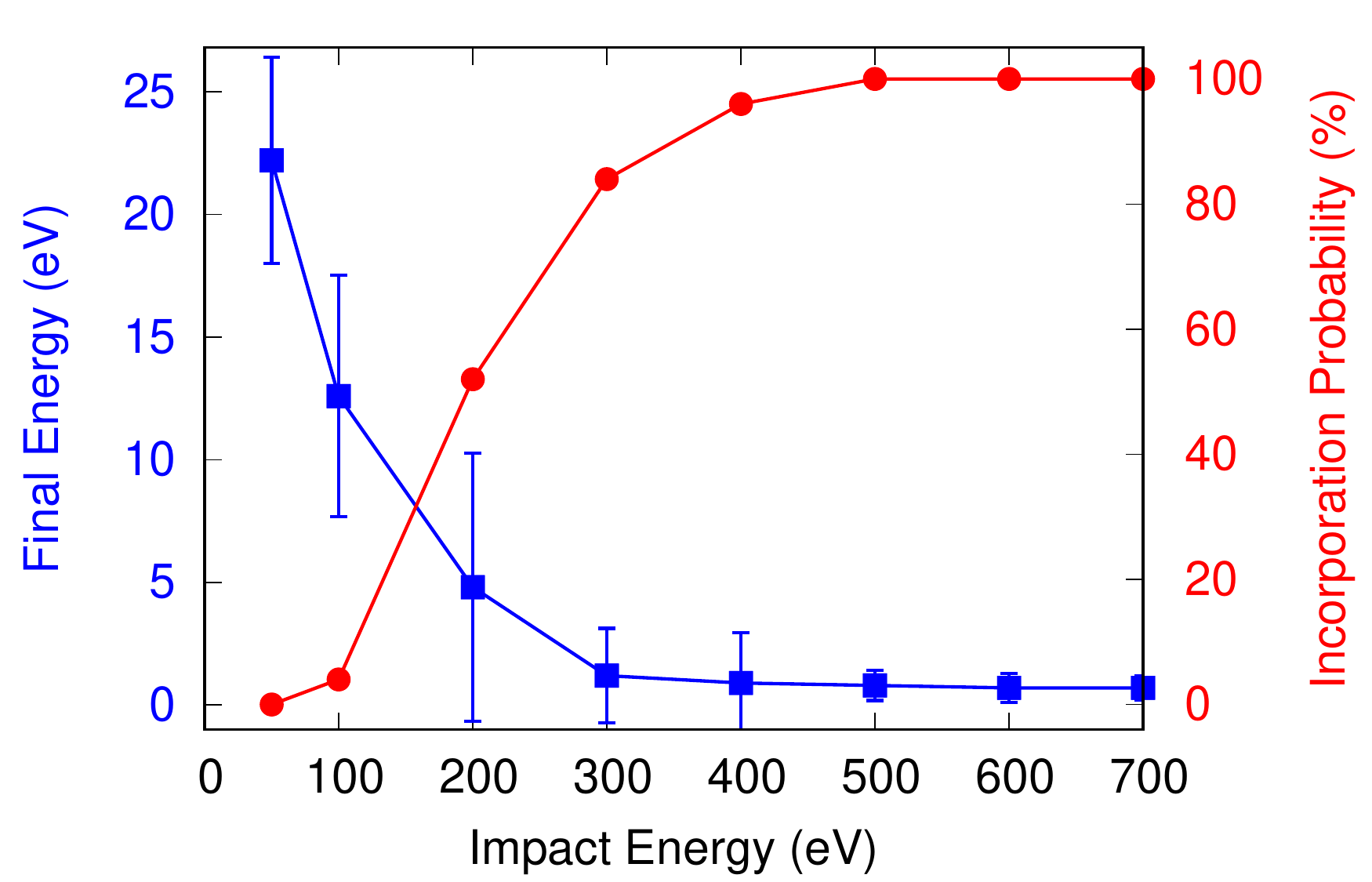}
\caption{ Final kinetic energy of the Xe atom (left-axis, blue squares), and implantation probability 
(right-axis, red circles) for 25 low-energy implantations onto a 3.1~nm ND. Error bars indicate one standard deviation. }
\label{low-energy}
\end{figure}

It is instructive to examine more closely the low-energy regime in
Fig.~\ref{PKAenergy-3.1nm}, since this is relevant to the ion-implantation
experiments \cite{Koscheev-Nature-2001,Verchovsky-Abs-2000} performed to
interpret Xe release from meteoritic NDs. Fig.~\ref{low-energy} shows that for
very low Xe energies (i.e.\ 50~eV) the Xe does not incorporate into the ND, and
is reflected with around half its initial kinetic energy. For an incident
energy of 200~eV there is a $\sim$50\% probability that the Xe will be
incorporated, increasing to 100\% at around 500~eV. For context, the
experiments by Koscheev et al. \cite{Koscheev-Nature-2001} used an implantation
energy of $\sim$700~eV while those of Verchovsky et al.
\cite{Verchovsky-Abs-2000} used $\sim$1000~eV and hence their experiments fall
in a regime where the Xe is always implanted at least several atomic layers
into the ND, but not with sufficient energy to pass through the ND. As a
secondary observation, the modest energies used in their experiments implies
only a small amount of heating as seen in Fig.~\ref{PKAenergy-3.1nm}; as a
result there will be no annealing effect such as occurs for implantations in
the vicinity of 6~keV. To clarify, these comments on the experimental
implantation studies apply only to a single impact, while the experiments used
doses of order $10^{15}$~ions/cm$^2$, corresponding to tens of ions per
square-nanometre. At these doses, individual NDs may be struck multiple times,
leading to accumulated damage beyond that considered in the present
simulations.

\begin{figure}[t]
\centering
\includegraphics[width=\columnwidth]{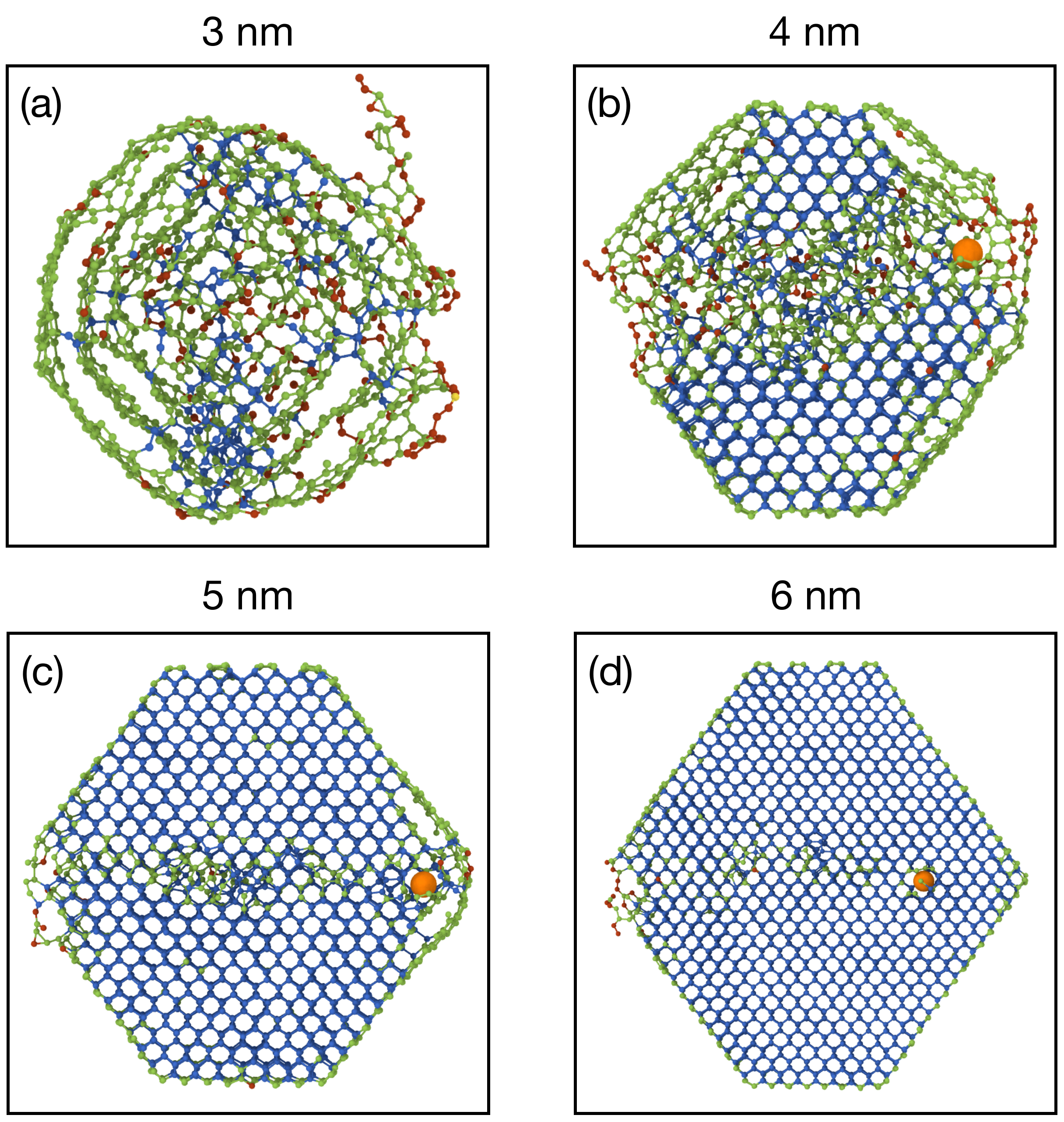}
\caption{ Cross-sectional snapshots (1~nm slices) showing how the size of the
ND influences damage following implantation of a 6~keV Xe from the left of
frame. All snapshots show the system 1~ps after impact.  Color
coding indicates the hybridization: $sp$, $sp^2$ and $sp^3$ atoms are 
red, green and blue circles, respectively. The Xe is shown as a large orange
circle. }
\label{6keV-snapshots}
\end{figure}

\begin{figure}[t]
\centering
\includegraphics[width=\columnwidth]{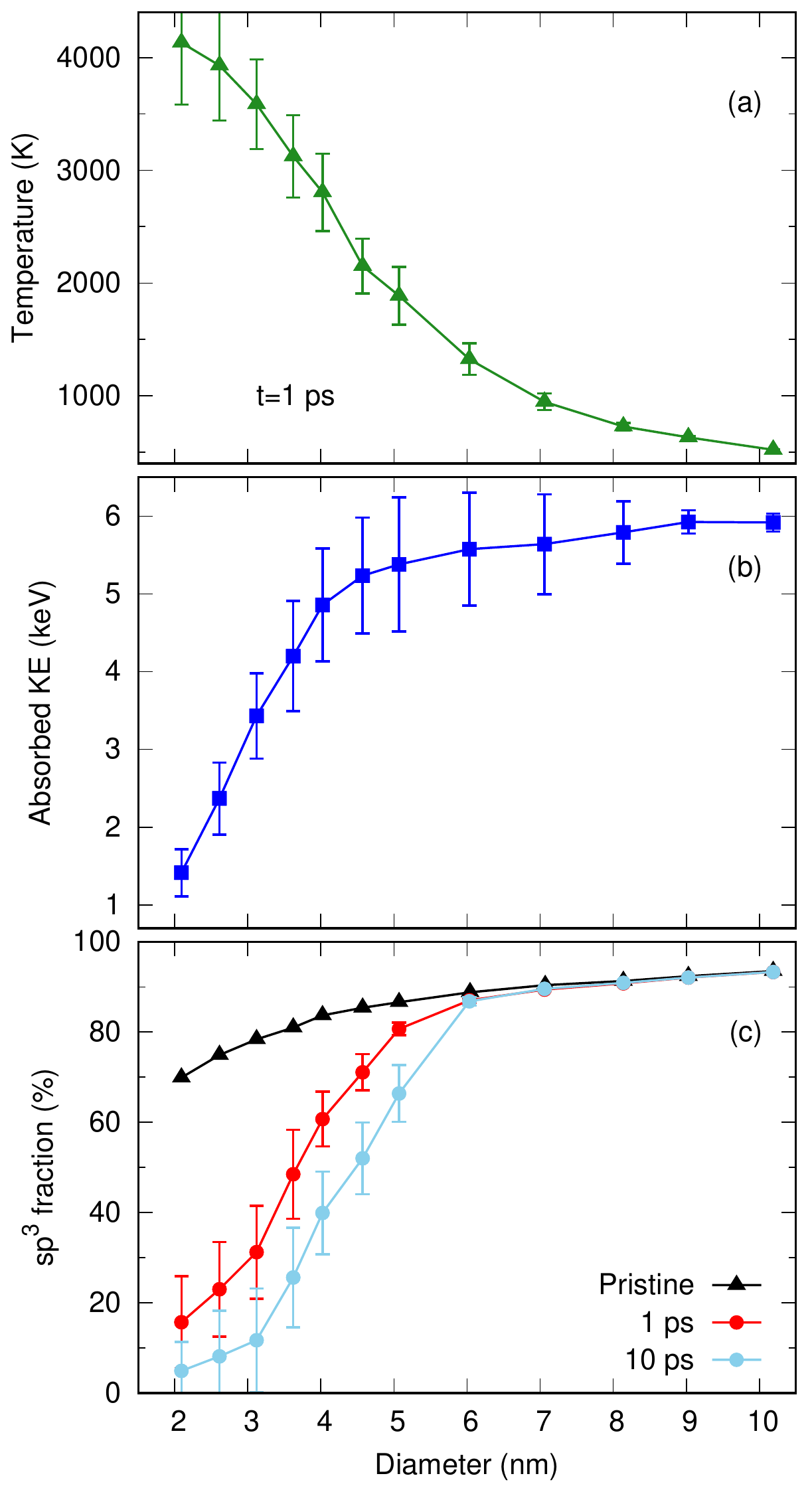}
\caption{ Effect of ND diameter for Xe implantation at a fixed PKA energy
of 6~keV.  All properties computed 1~ps after initiation of the PKA.
(a) Cluster temperature, (b) $sp^3$ fraction, and (c) kinetic energy
absorbed by the ND. }
\label{6keV-properties}
\end{figure}

All of the data discussed to this point has been for a 3.1~nm diameter ND.  To
study the effect of varying the ND size, a series of calculations were
performed using a 6~keV Xe and NDs ranging from 2.1 to 10.2~nm.
Cross-sectional snapshots in Fig.~\ref{6keV-snapshots} demonstrate that the
size of the ND strongly influences the nature of the damage created by the Xe
PKA. For larger diameters, such as the 5 and 6~nm NDs seen in panels (c,d),
there is minimal disruption to the ND itself, and damage is concentrated along
the trajectory of the Xe. In the case of the 6~nm ND, the residual damage
consists of isolated point defects, much the same as for cascades in bulk
diamond; see simulations by Buchan et al.~\cite{Buchan-JAP-2015}.  For the
4~nm ND the damage caused by the Xe is much more substantial, with a wide
damage track and graphitization on several (111) faces, while at 3~nm the ND is
largely transformed into a carbon onion.

The size-dependence effect seen in Fig.~\ref{6keV-snapshots} is explained by
Fig.~\ref{6keV-properties} which quantifies the temperature, $sp^3$ fraction
and kinetic energy absorbed by the ND. Panel (a) shows that the cluster
temperature falls off rapidly as the size of the ND increases; at 3~nm the
average temperature after 1~ps is above 3500~K, close to the melting point of
diamond. This high temperature explains the onionization behaviour seen in
Fig.~\ref{6keV-snapshots}(a). For a 4~nm ND the temperature is around 700~K
lower, reducing the extent of onionization but still sufficient to cause some
structural rearrangement as seen in Fig.~\ref{6keV-snapshots}(b).  For the 5
and 6~nm NDs the average temperature at 1~ps is circa 1900 and 1300~K,
respectively, far below the diamond melting point, explaining why the damage
tracks resemble those of bulk diamond. 

Fig.~\ref{6keV-properties}(b) quantifies the amount of kinetic energy which
the Xe PKA deposits into the ND. This quantity is determined by summing the
kinetic energy of all ejected particles (xenon plus carbons) and calculating
the difference from the original value of 6~keV. The graph shows that the small
NDs absorb only a small fraction of the total PKA energy, while for the large
NDs the entire energy of the PKA is deposited into the ND. Even though more
kinetic energy is absorbed by the large NDs, the final cluster temperature as
shown in Fig.~\ref{6keV-properties}(a) monotonically decreases with increasing
diameter due to the rapid increase in number of atoms with diameter. This
demonstrates an important aspect of ion implantation into nanoparticles,
namely that a small particle will only absorb a modest fraction of the PKA
energy, but since a small particle contains few atoms, the resultant
temperature increase is substantial.

\begin{figure*}[t]
\centering
\includegraphics[width=2.05\columnwidth]{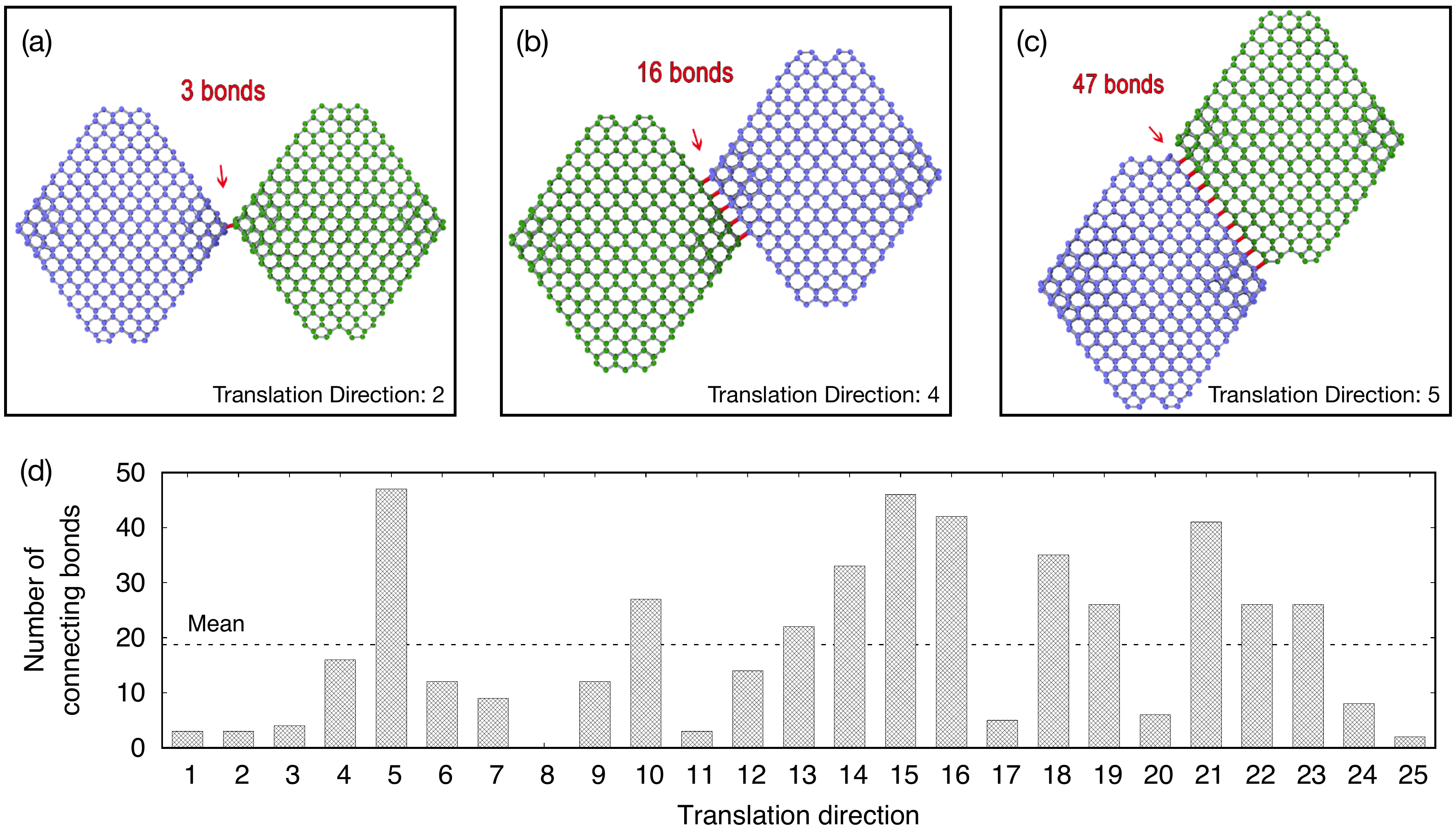}
\caption{(a--c) Cross-sectional snapshots of three of the double-ND systems
used to study the effect of thermal conduction. (d) Bar plot of the 25
different translation directions showing the number of bonds connecting
the two NDs. Panel (a) shows the weakest thermal contact, comprising just
3 bonds. Panel (b) shows typical (average) thermal contact, with 16  bonds. Panel (c) shows the 
highest thermal contact, with 47 bonds.
Connecting bonds are shown in red. Blue and green circles denote atoms and data 
for the primary and secondary ND, respectively.}
\label{2ND-connections}
\end{figure*}

Fig.~\ref{6keV-properties}(c) quantifies the fraction of $sp^3$ bonds as
a function of ND diameter; note that the maximum value (solid
triangles) varies with diameter due to the changing surface-to-volume
ratio. The loss of $sp^3$ bonding is particularly strong for the small
NDs. After 1~ps, around 16\% of the atoms in the 2~nm ND are $sp^3$
bonded, as compared to 70\% $sp^3$ bonding in the original ND. As the
diameter increases, the loss of $sp^3$ bonding gradually reduces, but 
remains significant in the 3--4~nm range.  For NDs around 6~nm and
above the sp$^3$ fraction is virtually unchanged due to the passage
of the Xe. 

The light-blue circles in Fig.~\ref{6keV-properties}(c) show the $sp^3$ fraction after 10~ps has 
elapsed. Since the NDs in these simulations are isolated, the residual thermal energy drives an 
annealing process which gradually graphitizes the smaller NDs. In the experimental situation however, 
thermal conduction with surrounding NDs will provide a heat-loss path, and hence the annealing 
process will not continue indefinitely. This raises the obvious question as to what simulation should 
be used; thus far we have used 1~ps, but if thermal conduction is very rapid then a shorter 
simulation time should be used, and likewise, if thermal conduction is poor, then the NDs should be 
allowed to self-anneal for longer. 

To address the question of thermal conduction between NDs, a second suite of
calculations were performed in a system containing two identical NDs in varying
degrees of thermal contact. The starting structure was the 3.1~nm ND, which was
replicated and displaced along a vector correspond to one of the 25 Thomson
points. The displacement was increased in small increments of 0.1~\AA\, until
the smallest bondlength in the entire double-ND system exceeded a threshold of
1.3~\AA. The structure was then relaxed using steepest descent minimization.
This process was repeated for the other 24 Thomson points, yielding a set of 25
double-ND systems exhibiting a wide range of thermal contact.  The system with
the weakest thermal contact is shown in Fig.~\ref{2ND-connections}(a),
containing just 3 connecting bonds (red lines) connecting the two NDs. In
contrast, the best thermal contact involved 47 connecting bonds as shown in
Fig.~\ref{2ND-connections}(c). To determine a typical value of connections, we
computed the number of connections for all 25 systems, as shown in the bar plot
of Fig.~\ref{2ND-connections}. This shows that the average number of
connections is around 18, for which the structure in
Fig.~\ref{2ND-connections}(b) is a representative example. 

\begin{figure*}[t]
\centering
\includegraphics[width=2\columnwidth]{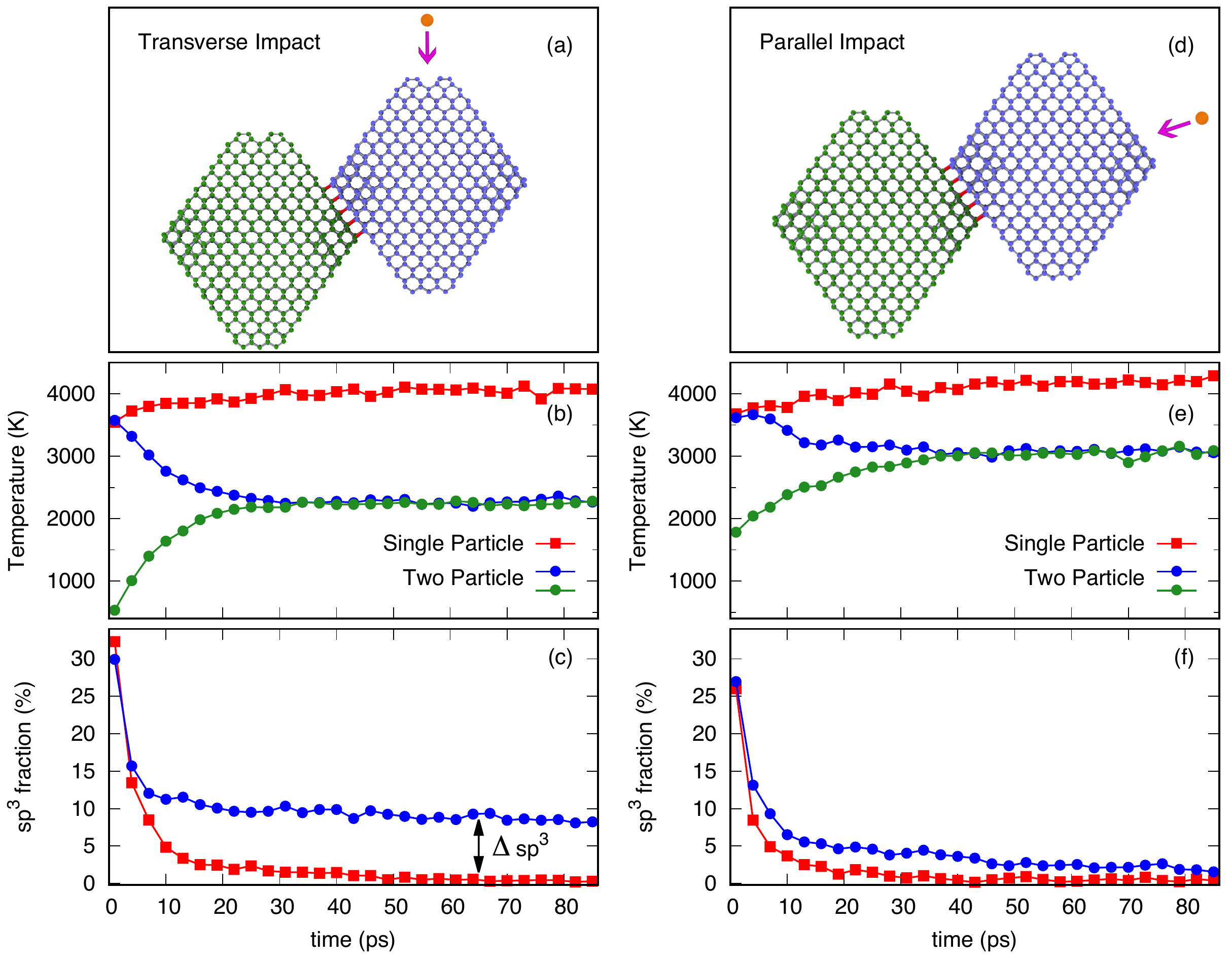}
\caption{ Time-evolution of temperature and $sp^3$ fraction for a 6~keV Xe impact into a double-ND 
system with average thermal contact [see Fig.~\ref{2ND-connections}(b)]. The first data point is 
1~ps after initiation of the PKA, followed by data in 3~ps intervals.
Panels (a--c) show the initial conditions and data for a transverse impact, in which the PKA 
direction is maximally orthogonal to the translation vector. Panels (d--f) show the same quantities 
for a parallel impact, in which the PKA direction is the same as the translation vector. Blue and 
green circles denote atoms and data for the primary and secondary ND, respectively. Red squares 
denote data for a single ND. The orange circle and pink arrow indicate the xenon atom and its initial 
direction, respectively. } 
\label{2ND-average}
\end{figure*}

For all 25 double-ND systems, two 6~keV PKA simulations were performed. These
simulations, which we refer to as either transverse or parallel, are shown in
detail for the case of average thermal contact in Fig.~\ref{2ND-average}. In
the transverse impact case, the PKA direction is the Thomson point vector which
is maximally orthogonal to the translation vector, while for the parallel
impact case the PKA direction is the same as the translation vector. The blue
and green circles in Fig.~\ref{2ND-average}(b) show the temperature of the
individual NDs as a function of time, using the same colour scheme employed in
panel (a) for the atoms. The heat-loss path provided by the secondary ND (green
circles) has an obvious effect on the temperature of the primary ND, shown in
blue. After approximately 30~ps the two NDs attain thermal equilibrium with
each other, reaching a temperature of $\sim$2250~K. The red squares in
Fig.~\ref{2ND-average}(b) show the exact same impact for a single-ND system in
which there is no heat-loss path. In this case, the temperature of the ND is
substantially higher, equilibrating at around 4000~K. The effect of this higher
temperature on the $sp^3$ fraction of the primary-ND is shown in
Fig.~\ref{2ND-average}(c), with red and blue circles corresponding to the
single- and double-ND systems, respectively. For the single-ND case, all $sp^3$
bonding is lost and the ND transforms into a carbon onion, while for the
double-ND around 8\% of the $sp^3$ bonds persist. While the effect of the
heat-loss path is clearly apparent, it is important to recall that the original
$sp^3$ fraction is 78\%, and hence in both cases substantial modification of
the ND occurs.

The parallel impact case shown in Fig.~\ref{2ND-average}(d--f) demonstrates
that the effect of thermal contact is influenced by the PKA direction. In this
instance, the xenon atom is directed towards the centre of mass of both NDs,
and this alignment means ballistic transfer of energy occurs between the
primary and secondary NDs. As a result, the secondary-ND is no longer just a
heatsink as it was in the transverse impact case. Panel (e) shows that the
equilibrium temperature is 3000~K, substantially higher than seen for the
transverse impact,and much closer to the single impact data (red squares). As
would be expected, the similarity in temperatures between the single- and
double-ND systems means that the difference $\Delta sp^3$ in panel (f) is
minimal. 

\begin{figure}[t]
\centering
\includegraphics[width=\columnwidth]{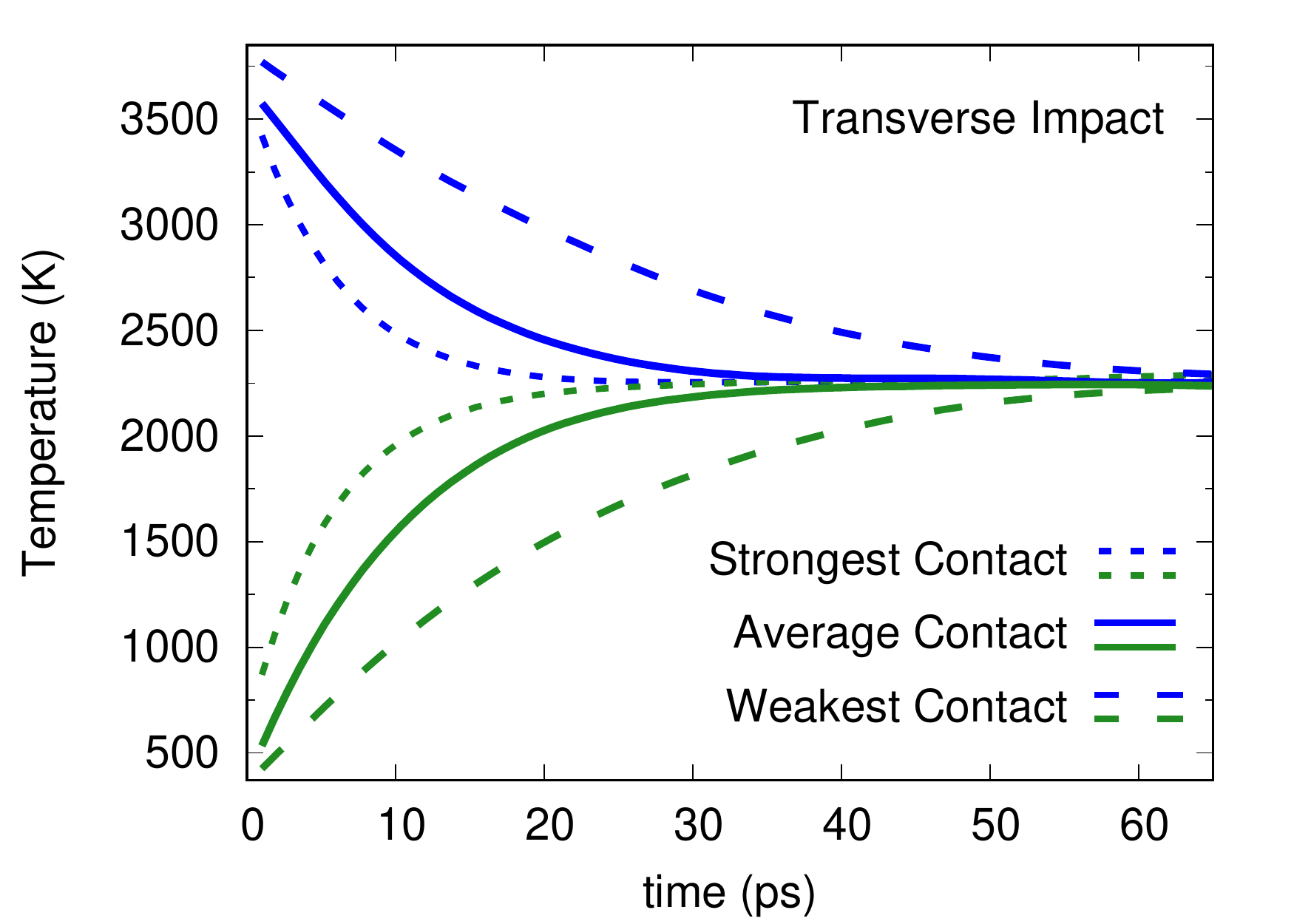}
\caption{Time-evolution of temperature for a 6~keV transverse impact onto the double-ND systems corresponding to weakest thermal contact [Fig.~\ref{2ND-connections}(a)], average thermal contact
[Fig.~\ref{2ND-connections}(b)] and strongest thermal contact [Fig.~\ref{2ND-connections}(c)]. The blue and green traces are decaying exponential fits to the raw data, and correspond to the primary and secondary ND, respectively. }
\label{2ND-exponential}
\end{figure}

\begin{figure}[b]
\centering
\includegraphics[width=\columnwidth]{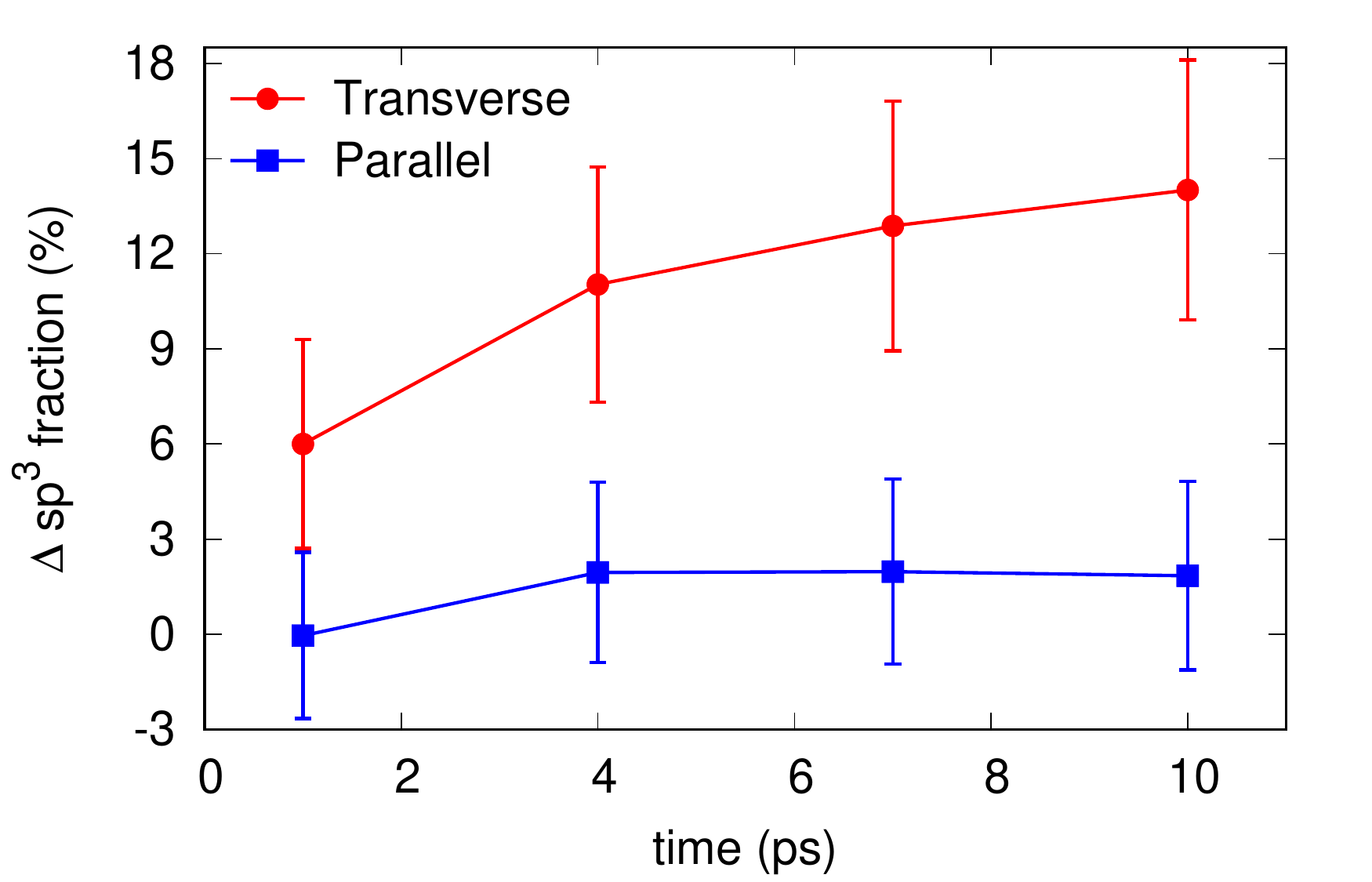}
\caption{Effect of annealing time on the difference in $sp^3$ fraction between 
single- and double-ND systems for transverse (red circles) and parallel (blue 
squares) impacts. The sign convention is defined so that a positive difference means 
the double-ND value is higher. Data collected for all 25 of the translation 
directions (see Fig.~\ref{2ND-connections}) and the error bars indicate one 
standard deviation. }
\label{2ND-sp3}
\end{figure}

The equilibration between the primary and secondary ND seen in the
non-ballistic, transverse impact data of Fig.~\ref{2ND-average}(b) can be
intuitively understood from a macroscopic perspective using Newton's law of
heating/cooling. Both temperature profiles are well-fitted by decaying
exponentials, with exponents of 9.0 and 8.4~ps$^{-1}$ for the primary and
secondary ND, respectively. The solid lines in Fig.~\ref{2ND-exponential} shows
the corresponding fitted exponential curves, along with data for the cases of
strongest and weakest thermal contact. The average decay constant for the
strongest contact is 4.2~ps$^{-1}$, while for the weakest contact case it is
20~ps$^{-1}$. This data provides a timescale for the heat-loss path when NDs
are in contact with one another. This situation is directly relevant to the
\emph{in situ} implantation/TEM experiments which motivated the present
simulations. In the experiments, NDs were suspended in ethanol and deposited
onto a holey carbon grid, producing an aggregate of NDs (see Fig. 1(a) in
Ref.~\cite{Shiryaev-SciRep-2018}) that provide a heat-loss path. Based on the
simulation results, it is reasonable to propose that 1~ps of annealing is about
right, while 10~ps is certainly too much, particularly since each ND will have
some degree of thermal contact with several other NDs. 

The effect of having too much annealing is illustrated in Fig.~\ref{2ND-sp3}
which plots the quantity $\Delta sp^3$ (as defined in Fig.~\ref{2ND-average})
as function of annealing time for transverse and parallel impacts. These two
situations can be considered as the extrema, with a typical impact falling
somewhere in-between. All data points are averages across 25 different
translation directions and the errors denote one standard deviation. This
sample of is sufficient size to display a clear trend in which overly long
annealing times lead to significantly different $sp^3$ fractions for transverse
impacts. In contrast, there is little effect for parallel impacts as the
kinetic energy of the atoms ejected from the primary-ND is deposited into the
secondary-ND; this increases the temperature of the secondary-ND, thereby
reducing the temperature gradient along the heat-loss path.   

\section{Conclusion}

In this paper, we showed how molecular dynamics simulation can be quantatively
used to study implantation of xenon into nanodiamonds (NDs). Following on from
our previous study \cite{Shiryaev-SciRep-2018}, we provided detail on the key
experimental/simulation result, namely, that a primary knock-on atom (PKA)
energy around 6~keV provides maximal damage.  We also detailed a pronounced
size effect, showing that small NDs (below 3--4~nm diameter) are easily
destroyed, losing many atoms and graphitizing those which remain. An important
aspect of the damage process is the interplay between the amount of kinetic
energy deposited in the ND, and the number of atoms in the ND itself.  This
energy transfer leads to thermal heating of the ND, with an effect seen only
when the temperature approaches the rather high melting point of diamond.  To
assess the length of annealing that should be used in the simulations, we also
studied a double-ND system in which two particles are touching with varying
degrees of thermal contact. This mimics the heat bath that occurs in the
experiment, and suggests an annealing time of one picosecond is reasonable.

The methodology developed here can be easily extended to describe a variety of
similar topics. The most obvious is the implantation of noble gases into NDs,
which is relevant to presolar studies where all of the noble gases are
important. To extend the present work to elements other than xenon one simply
needs to take Lennard-Jones parameters for noble gas interactions with carbon
and bridge to the short-range ZBL expression as in Fig.~\ref{potential}. These
parameters are easily obtainable due to the many studies of gas-adsorption onto
carbon networks, and are conceptually straightforward due to the weak
interactions between carbon and a noble gas atoms. Implantation involving other
chemical species is also possible, but the interactions are less trivial due to
the formation of formal chemical bonds.  Beyond NDs, the noble gas implantation
methodology can also be usefully applied to study ion-irradiation effects in a
wide variety of different carbon nanoforms, such as graphene, graphene
bilayers, fullerenes and nanotubes (see Ref.~\cite{Krasheninnikov-JAP-2010} for
a summary of this field), as well as bulk materials such as glassy carbon,
graphite and diamond.

\section*{Acknowledgements}

This work was supported by Australian Research Council via Project FT120100924. Computational 
resources were provided by the Pawsey Supercomputing Centre with funding from the Australian 
Government and the Government of Western Australia.

\bibliography{references}

\end{document}